\documentclass[%
 aip,
 jmp,%
 amsmath,amssymb,
 reprint,%
]{revtex4-1}

\usepackage{graphicx}
\usepackage{dcolumn}
\usepackage{bm}

\begin{document}

\preprint{AIP/123-QED}

\title{Velocity relaxation of a particle in a confined compressible fluid}

\author{Rei Tatsumi}
 \email{tatsumi@cheme.kyoto-u.ac.jp}
\author{Ryoichi Yamamoto}%
 \email{ryoichi@cheme.kyoto-u.ac.jp}
\affiliation{%
 Department of Chemical Engineering, Kyoto University, Kyoto 615-8510, Japan
}%


\date{\today}

\begin{abstract}
The velocity relaxation of an impulsively forced spherical particle in a fluid confined by two parallel plane walls is studied using a direct numerical simulation approach.
During the relaxation process, the momentum of the particle is transmitted in the ambient fluid by viscous diffusion and sound wave propagation, 
and the fluid flow accompanied by each mechanism has a different character and affects the particle motion differently.
Because of the bounding walls, viscous diffusion is hampered, and the accompanying shear flow is gradually diminished. 
However, the sound wave is repeatedly reflected and spreads diffusely.
As a result, the particle motion is governed by the sound wave and backtracks differently in a bulk fluid.
The time when the backtracking of the particle occurs changes non-monotonically with respect to the compressibility factor $\varepsilon = \nu /ac$ 
and is minimized at the characteristic compressibility factor.
This factor depends on the wall spacing, 
and the dependence is different at small and large wall spacing regions based on the different mechanisms causing the backtracking.
\end{abstract}

\pacs{Valid PACS appear here}
\maketitle

%

\section{\label{sec:level1}Introduction}
The dynamics of particles dispersed in a fluid flowing through a microchannel are important for many chemical engineering processes, such as membrane separation and microfluidics.
The dynamics of fluids and dispersed particles are significantly affected by bounding walls; thus, 
the transport properties of dispersions in some simple bounding geometries have been studied using analytical methods~\cite{B4-1}.
While such studies were limited to a steady flow system, 
recently, the unsteady dynamics of a dispersed particle in a confined fluid have been investigated~\cite{B4-2,B4-3,B4-4,B4-5,B4-6,B4-7,B4-8}.

In a dispersion system, the momentum of a particle is propagated in the ambient fluid via two mechanisms: 
viscous diffusion and sound propagation. 
Each of these mechanisms is accompanied by fluid flow of different character, which affects the particle motion.
Here, we consider the velocity relaxation of a particle after adding an impulsive force. 
In a bulk fluid, part of the particle momentum is transported by a sound wave to an infinite distance in time, 
and finally, the particle motion is governed by shear flow accompanied by viscous diffusion, 
which results in the algebraic decay obeying $t^{-3/2}$~\cite{B4-9,B4-10}.
However, in a fluid confined by walls with stick boundary conditions, 
both the viscous diffusion and sound propagation are affected by the walls.
The viscous diffusion is hampered at the walls, and the accompanying shear flow gradually diminishes; 
however, the sound wave is repeatedly reflected at the walls and spreads diffusely~\cite{B4-3,B4-4}.
Consequently, the particle motion is finally governed by flow accompanied by the spreading sound wave.
Especially in a fluid confined between two parallel plane walls,
the particle velocity relaxation presents a negative $t^{-2}$ long-time decay differently than that presented in a bulk fluid, 
which is derived by the mode-coupling theory~\cite{B4-3} and Green's function method~\cite{B4-4}.

In the present study, we investigate the dynamics of a single spherical particle 
in a fluid confined by two parallel plane walls with stick boundary conditions using a direct numerical simulation approach.
We use the smoothed profile method (SPM)~\cite{B4-11,B4-12}, which is applicable to a compressible fluid~\cite{B4-13,B4-13a}.
In SPM, rigid fixed wall boundaries can be imposed similar to the representation of rigid particles~\cite{B4-14}.
The accuracy of SPM for the present system is confirmed by calculating the steady-state mobility of the particle, 
which is compared with the approximate analytical solutions.
We examine the velocity relaxation of the particle after adding an impulsive force in the direction parallel to the walls.
The velocity relaxation function corresponds to the velocity autocorrelation function in a fluctuating system.
We first consider an incompressible fluid to investigate the wall effects on the dynamics governed only by viscous diffusion. 
We then consider a compressible fluid and investigate the particle motion affected by 
the spreading sound wave arising from reflections at the walls.

\section{Model}

\begin{figure}[tbp]
\centering
\includegraphics[width=85mm, clip]{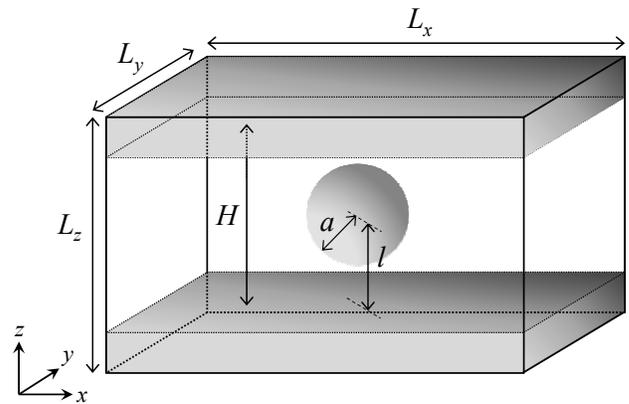}\\[-0.7em]
\caption{\label{f4-1} Geometry of the present model system.
A single spherical particle is in a fluid confined by two parallel plane walls.
The particle radius is $a$, and the wall spacing is $H$.
The particle is located at a height $l$ from the lower wall.
The system is considered to be in a simulation box of size $L_x \times L_y \times L_z$.
The periodic boundary conditions are imposed on the end of the box in all directions.
The walls are set on the top and bottom of the box in the $z$ direction with thickness $(L_z-H)/2$.
The walls generate anisotropy for two directions: parallel and perpendicular to the walls.
In this figure, the $x$ and $y$ directions are the degenerate parallel directions, 
and the $z$ direction is the perpendicular direction.
}
\end{figure}

We consider a system in which a single particle is dispersed in a Newtonian fluid confined by 
two parallel plane walls, as described in Fig.~\ref{f4-1}.
The stick boundary conditions are imposed on the walls.
The motion of the particle is governed by Newton's and Euler's equations of motion as
\begin{eqnarray}
M \frac{\mathrm{d}}{\mathrm{d} t} \boldsymbol{V} = \boldsymbol{F}^H + \boldsymbol{F}^W + \boldsymbol{F}^E,\ \ \ \ 
\frac{\mathrm{d}}{\mathrm{d} t} \boldsymbol{R} = \boldsymbol{V}
\label{e4-2-1},
\end{eqnarray}
\begin{eqnarray}
\boldsymbol{I} \cdot \frac{\mathrm{d}}{\mathrm{d} t} \boldsymbol{\Omega} = \boldsymbol{N}^H + \boldsymbol{N}^E
\label{e4-2-2},
\end{eqnarray}
where $\boldsymbol{R}$, $\boldsymbol{V}$, and $\boldsymbol{\Omega}$ are the position, translational velocity, 
and rotational velocity of the particle, respectively.
The particle has a mass $M$ and a moment of inertia $\boldsymbol{I}$.
A hydrodynamic force $\boldsymbol{F}^H$ and a torque $\boldsymbol{N}^H$ are exerted on the particle by the fluid,
and a repulsive force $\boldsymbol{F}^W$ prevents the particle from overlapping with the walls.
A force $\boldsymbol{F}^E$ and a torque $\boldsymbol{N}^E$ are externally applied.
The hydrodynamic force and torque are evaluated by simultaneously considering the fluid motion.

The dynamics of fluid are governed by the following hydrodynamic equations:
\begin{eqnarray}
\frac{\partial \rho}{\partial t} + \boldsymbol{\nabla} \cdot (\rho \boldsymbol{v}) = 0
\label{e4-2-3},
\end{eqnarray}
\begin{eqnarray}
\frac{\partial \rho \boldsymbol{v}}{\partial t} + \boldsymbol{\nabla} \cdot (\rho \boldsymbol{vv}) 
= \boldsymbol{\nabla} \cdot \boldsymbol{\sigma} + \rho \phi_P \boldsymbol{f}_P + \rho \phi_W \boldsymbol{f}_W
\label{e4-2-4},
\end{eqnarray}
where $\rho(\boldsymbol{r},t)$ and $\boldsymbol{v}(\boldsymbol{r},t)$ are the mass density and velocity fields of the fluid, respectively.
The stress tensor is given by
\begin{eqnarray}
\boldsymbol{\sigma} = -p \boldsymbol{I} + \eta [\boldsymbol{\nabla} \boldsymbol{v} + (\boldsymbol{\nabla} \boldsymbol{v})^T]
 + \left( \eta_v - \frac{2}{3}\eta \right)(\boldsymbol{\nabla} \cdot \boldsymbol{v}) \boldsymbol{I}, \nonumber \\
\label{e4-2-5}
\end{eqnarray}
where $p(\boldsymbol{r},t)$ is the pressure, $\eta$ is the shear viscosity, and $\eta_v$ is the bulk viscosity.
The body force $\rho \phi_P \boldsymbol{f}_P$ is added such that the rigidity of the particles is satisfied.
The external force field $\rho \phi_W \boldsymbol{f}_W$ is also introduced 
to impose the stick boundary conditions by two parallel plane walls 
such that the force $\rho \phi_W \boldsymbol{f}_W$ constrains the velocity field in the wall region to be zero.
Additionally, we assume a barotropic fluid described by $p = p(\rho)$,
with the constant speed of sound $c$ being
\begin{eqnarray}
\frac{\mathrm{d} p}{\mathrm{d} \rho} = c^2
\label{e4-2-6}.
\end{eqnarray}
Equations~(\ref{e4-2-3})-(\ref{e4-2-6}) are closed for the variables $\rho$, $\boldsymbol{v}$, and $p$ 
without consideration of energy conservation.

We use the SPM for the direct numerical simulations in the present study.
The system is composed of three regions: the fluid, particle, and wall. 
In SPM, the boundaries between the fluid region and the other regions are expressed by the continuous phase profile function.
There are no boundaries between the particle and wall regions because the particle does not penetrate into the wall.
For this purpose, we introduce a smoothed profile function $\phi_X(\boldsymbol{r}, t) \in [0, 1]$, 
where the index $X$ signifies the region of the particle $P$ or wall $W$.
The function $\phi_X$ represents the boundary between the regions of the fluid and $X$, 
such that $\phi_X = 1$ in region $X$ and $\phi_X = 0$ in the other regions. 
Using the smoothed profile function, the regions of the fluid and $X$ are smoothly connected through a thin interfacial region with thickness $\xi$.
The detailed mathematical expression of $\phi_X$ is given in a previous paper~\cite{B4-11}.

\section{Numerical Results}

Numerical simulations are performed for a three-dimensional box with periodic boundary conditions.
The space is divided into meshes of length $\Delta$, 
which is a unit length.
The units of the other physical quantities are defined by combining $\eta = 1$ and $\rho_0 = 1$ with $\Delta$, 
where $\rho_0$ is the fluid mass density at equilibrium.
The other parameters are set to $a = 4$, $\xi = 2$, $\rho_p = 1$, $\eta_v = 0$, and $h = 0.05$,
where $\rho_p$ is the particle mass density, and $h$ is the time increment of a single simulation step.
The geometry of the present system is described in Fig.~\ref{f4-1}.
The particle is set on the midway between two walls as $l = H/2$.
The ratio of the wall spacing to the particle radius, $H^{\ast} = H/a$, is set to various values.
The system size is $L_x \times L_y \times L_z = 512 \times 512 \times 32$ for $H^{\ast} < 8$
and $L_z = 64$ for $H^{\ast} \geq 8$.

We investigate the relaxation of the particle velocity after exerting an impulsive force at the center of the particle. 
The impulsive force is assumed to be sufficiently small such that the Reynolds and Mach numbers of the flow are sufficiently low. 
We set the impulsive force to produce an initial particle Reynolds number of ${\rm Re}_p = 10^{-3}$.
In the considered system, two directions are specified: parallel and perpendicular to the walls;
therefore, assuming a low Reynolds number flow, the relaxation of the particle velocity is described as
\begin{eqnarray}
\boldsymbol{V}(t) = \frac{\boldsymbol{P}}{M} \cdot \boldsymbol{\gamma}(t)
\label{e4-3-1},
\end{eqnarray}
\begin{eqnarray}
\boldsymbol{\gamma}(t) = \gamma^{\parallel}(t)(\boldsymbol{I} - \hat{\boldsymbol{z}}\hat{\boldsymbol{z}}) 
+ \gamma^{\perp}(t)\hat{\boldsymbol{z}}\hat{\boldsymbol{z}}
\label{e4-3-2},
\end{eqnarray}
where $\boldsymbol{P}$ is the impulsive force exerted on the particle at $t = 0$ 
and $\hat{\boldsymbol{z}}$ is the unit vector in the $z$ direction.
The velocity relaxation tensor $\boldsymbol{\gamma}(t)$ also depends on the wall spacing $H$ and the distance of the particle from the lower wall $l$.
In the present study, we focus on the parallel motion of the particle $\gamma^{\parallel}(t)$;
therefore, in the following section, we represent the velocity relaxation function in the parallel direction by $\gamma(t)$ for simplicity.
For this reason, collisions of the particle against the wall will not occur, 
and the direct interactions between a particle and wall, including the overlap repulsion force, are not considered in the present simulations.
According to the fluctuation-dissipation theorem, 
the velocity relaxation function is equivalent to the velocity autocorrelation function in a fluctuation system:
\begin{eqnarray}
\boldsymbol{\gamma}(t) = 
\frac{M}{k_B T} \langle \boldsymbol{V}(0) \boldsymbol{V}(t) \rangle
\label{e4-3-3},
\end{eqnarray}
where $k_B$ is the Boltzmann constant, and $T$ is the thermodynamic temperature.

The important time scales in the dynamics of a single particle are 
those of viscous diffusion and sound propagation over the length of the particle radius: 
$\tau_\nu = a^2/\nu$ and $\tau_c = a/c$, respectively, where $\nu = \eta/\rho_0$ is the kinematic viscosity. 
We define the compressibility factor by the ratio of these time scales as
\begin{eqnarray}
\varepsilon = \frac{\tau_c}{\tau_\nu} = \frac{\nu}{ac}
\label{e4-1-1},
\end{eqnarray}
which provides a measure of the importance of sound propagation in the dynamics of a single particle.
In the present simulations, we adjust the fluid compressibility by the compressibility factor.

\subsection{Steady-state mobility in an incompressible fluid}

\begin{figure}[tb]
\centering
\includegraphics[width=90mm]{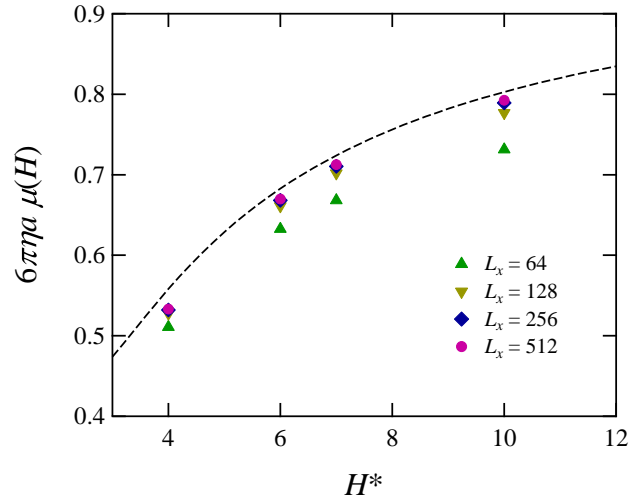}\\[-1.5em]
\caption{\label{f4-2} Wall spacing dependence of the mobility of a single particle in a fluid confined by two parallel plane walls.
The length of the simulation box sides parallel to the walls are changed to $L_x = L_y = 64$, 128, 256, and 512 to investigate the effect of the periodic boundary condition.
The broken line represents the analytical solution given by Eq.~(\ref{e4-3A-1}).
}
\end{figure}

First, we estimate the particle mobility $\mu$ in steady state for accuracy testing.
In these simulations, a constant force was continuously exerted on the particle in the direction parallel to the walls. 
The mobility was calculated as the ratio of the particle velocity to the added force after reaching steady state.
The mobilities in such situations have been studied by Fax{\'e}n using analytical theories~\cite{B4-1,B4-15}. 
The solution was given as the power series of the ratio of the wall spacing to particle radius, 
in which the first few terms were derived as
\begin{eqnarray}
6 \pi \eta a \mu(H) = 1 - 1.004 \lambda + 0.418 \lambda^3 \hspace{5.0em} \nonumber\\
+ 0.21 \lambda^4 - 0.169 \lambda^5 + O(\lambda^6)
\label{e4-3A-1},
\end{eqnarray}
where $\lambda = 2/H^{\ast}$.
The simulation results are presented in Fig.~\ref{f4-2}
and are compared with the approximate solutions given by Eq.~(\ref{e4-3A-1}).
The simulations were performed with the various system side lengths in parallel directions to the walls, $L_x$ and $L_y$ ($L_x = L_y$),
to investigate the system size effects due to the periodic boundary condition, 
which are diminished to an insignificant level at $L_x \geq 256$. 
However, the simulation results underestimate the mobility even at $L_x = 512$;
the deviations from the solutions given by Eq.~(\ref{e4-3A-1}) are less than $2 \%$ for $H^{\ast} \geq 6$ and approximately $5 \%$ only for $H^{\ast} = 4$.
As for $H^{\ast} = 4$, however, because the order of the power $\lambda^6$ is still 1, 
a truncation error of a few percent arises in Eq.~(\ref{e4-3A-1}).
Therefore, the simulation error for $H^{\ast} = 4$ can be less than $5 \%$.

From the present results, the walls with stick boundary conditions are successfully implemented by SPM.
The implementation of confinement by different geometries is also available in SPM.

\subsection{Velocity relaxation in an incompressible fluid}

\begin{figure}[tb]
\centering
\includegraphics[width=90mm]{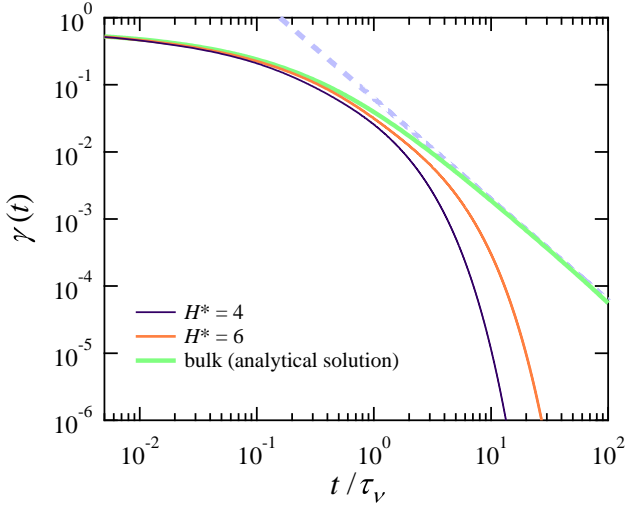}\\[-1.5em]
\caption{\label{f4-3} Velocity relaxation functions of a particle in an incompressible fluid confined by two parallel plane walls.
The wall spacings are $H^{\ast} = 4$ and 6.
The analytical solution of the corresponding function in a bulk fluid is also presented~\cite{B4-16}.
The bold dotted line represents the long-time tail $At^{-3/2}$, as given by Eq.~(\ref{e4-3B-2}).
}
\end{figure}

\begin{figure}[tbp]
\centering
\includegraphics[width=85mm, clip]{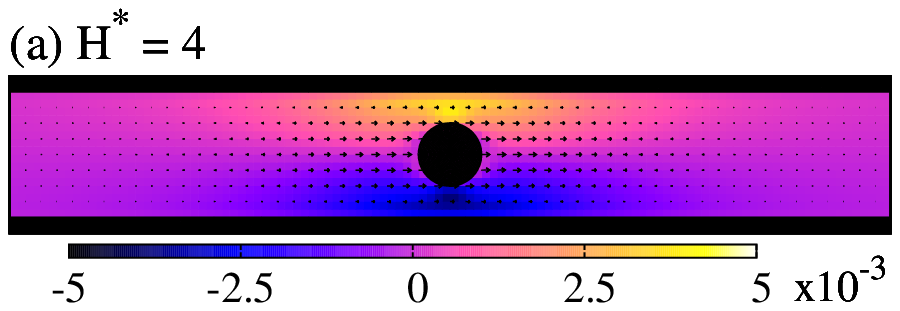}\\[-0.2em]
\includegraphics[width=85mm, clip]{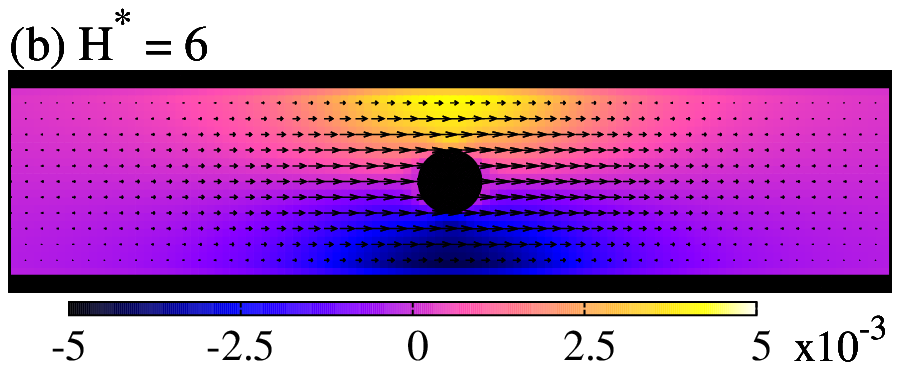}\\[-0.2em]
\includegraphics[width=85mm, clip]{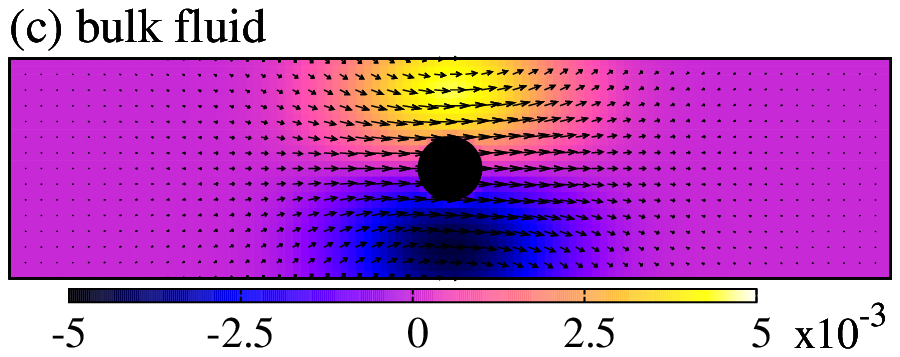}\\[-0.5em]
\caption{\label{f4-4} Velocity fields around the particle at $t/\tau_{\nu} = 2.81$ 
in confined fluids with a wall spacing of (a) $H^{\ast} = 4$ and (b) $H^{\ast} = 6$ and (c) in a bulk fluid are shown.
The walls are set on the upper and lower boundaries of the images for the confined fluids, 
and the corresponding region is presented for the bulk fluid.
The cross-sections presented here are perpendicular to the walls and parallel to the impulsive force direction and include the particle center.
The direction of the impulsive force is to the right in the images, 
and the particle is represented by a black circle.
The vorticity of the velocity field $\boldsymbol{\nabla} \times \boldsymbol{v}$ is described using a color scale, 
moving from negative (darker) to positive (lighter) vorticity.
The vorticity is normalized by the factor of $\tau_{\nu}/{\rm Re}_p$.
}
\end{figure}

Here, we consider an incompressible fluid as a solvent fluid. 
To treat an incompressible fluid, we assume an infinite speed of sound in Eq.~(\ref{e4-2-6}) and ignore the deviation of the fluid density.
With this assumption and the mass conservation law Eq.~(\ref{e4-2-3}),
the solenoidal condition for the velocity field is derived as
\begin{eqnarray}
\boldsymbol{\nabla} \cdot \boldsymbol{v} = 0
\label{e4-3B-1}.
\end{eqnarray}
Therefore, Eqs.~(\ref{e4-2-4}), (\ref{e4-2-5}), and (\ref{e4-3B-1}) were solved in combination as the hydrodynamic equations.

The simulation results of the velocity relaxation functions for wall spacings $H^{\ast} = 4$ and 6 are presented in Fig.~\ref{f4-3}.
The relaxation functions decrease monotonically, and the reduction rate increases with a decrease in the wall spacing $H^{\ast}$.
Exponential long-time decay is observed, as derived from the analytical theories~\cite{B4-17}.
Such decay contrasts the relaxation function in a bulk fluid, 
which describes long-time decay obeying the power law given by~\cite{B4-10}
\begin{eqnarray}
\gamma^{\rm bulk}(t) = \frac{1}{9 \sqrt{\pi}} \frac{\rho_p}{\rho_0} \left( \frac{\tau_{\nu}}{t} \right)^{3/2} \ \ \ {\rm as} \ \ t \rightarrow \infty
\label{e4-3B-2}.
\end{eqnarray}
The exponential decay of the relaxation function is expected to reflect the loss of fluid momentum at the walls with stick boundary conditions.
Therefore, the time when the relaxation function in a confined fluid starts to remarkably fall below that in a bulk fluid is 
related to the time scale when the fluid flow generated by the particle reaches the walls.
In an incompressible fluid, the temporal evolution of the fluid flow is only accompanied by viscous diffusion, 
whose time scale over the distance between the particle surface and the wall $(H/2-a)$ is 
given by $\tau_{\nu}^{\dagger} = (H/2-a)^2/\nu = (H^{\ast}/2-1)^2 \tau_{\nu}$.
For the wall spacings $H^{\ast} = 4$ and 6, the viscous diffusion time scales are $\tau_{\nu}^{\dagger}/\tau_{\nu} = 1$ and 4, respectively.
As demonstrated in Fig.~\ref{f4-3}, 
the starting time of the deviation of the function from that in a bulk fluid corresponds to the time scales $\tau_{\nu}^{\dagger}$.

The velocity fields around the particle with various wall spacings at the time $t/\tau_{\nu} = 2.81$ are displayed in Fig.~\ref{f4-4}.
Compared with a bulk fluid, the attenuation of fluid velocity is clearly observed in a confined fluid with $H^{\ast} = 4$, 
corresponding to the decay of the shear flow or the vorticity by the walls.
However, the attenuation of fluid velocity is not observed for $H^{\ast} = 6$.
This result coincides with the fact that the time $t/\tau_{\nu} = 2.81$ is earlier than the viscous diffusion time scale for $H^{\ast} = 6$, 
namely, $\tau_{\nu}^{\dagger}/\tau_{\nu} = 4$.
The walls restrict the viscous diffusion of the flow field to form a laminar flow, 
which corresponds to the extended distribution of the vorticity along the walls.

\subsection{Velocity relaxation in a compressible fluid}

\begin{figure}[tbp]
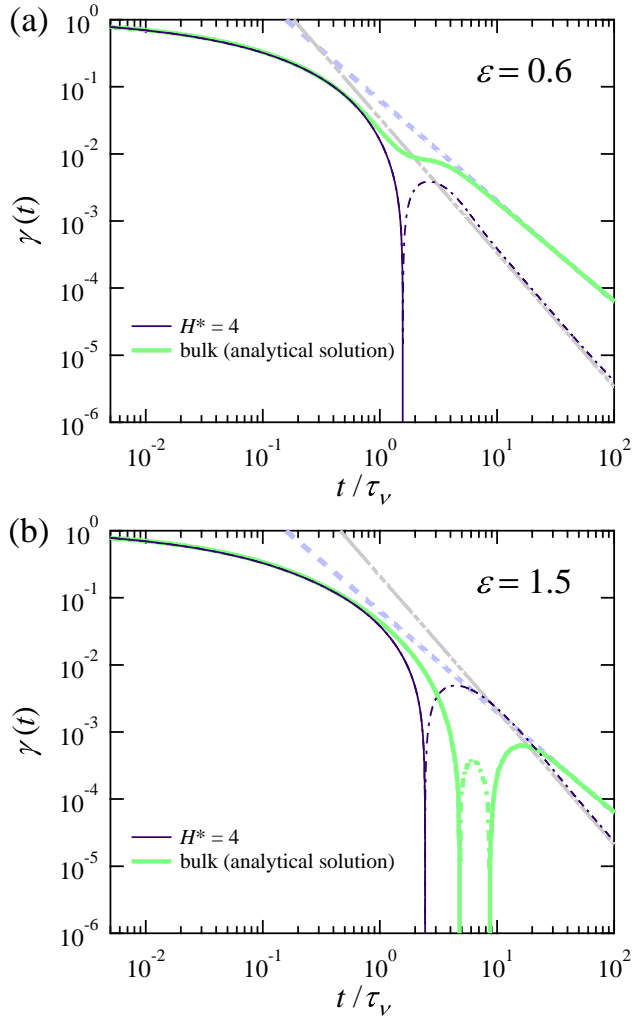

\centering
\includegraphics[width=90mm]{fig4_5a.eps}\\[-1.8em]
\includegraphics[width=90mm]{fig4_5b.eps}\\[-1.5em]
\caption{\label{f4-5} Velocity relaxation functions of a particle in a compressible fluid confined by two parallel plane walls with the wall spacing $H^{\ast} = 4$.
The compressibility factors are (a) $\varepsilon = 0.6$ and (b) $\varepsilon = 1.5$.
The analytical solutions of the corresponding functions in a bulk fluid are also presented~\cite{B4-18,B4-19}.
The functions are described by solid lines (positive values) and dashed-dotted lines (negative values).
The bold dotted lines represent the positive long-time tail $At^{-3/2}$, given by Eq.~(\ref{e4-3B-2}).
The bold dashed-two dotted lines represent the negative long-time tail $-Bt^{-2}$, given by Eq.~(\ref{e4-3C-1}).
}
\end{figure}

\begin{figure}[tbp]
\centering
\includegraphics[width=85mm, clip]{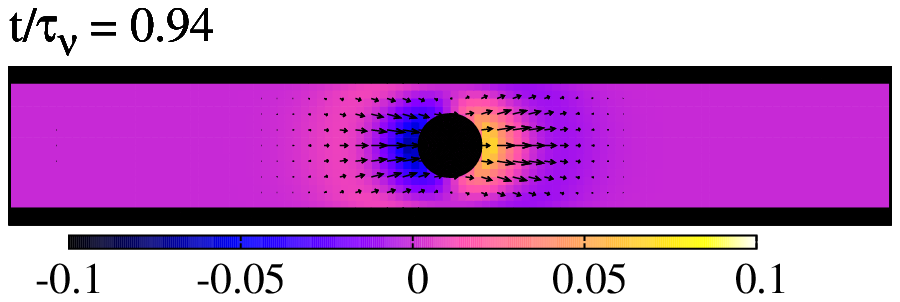}\\[-0.em]
\includegraphics[width=85mm, clip]{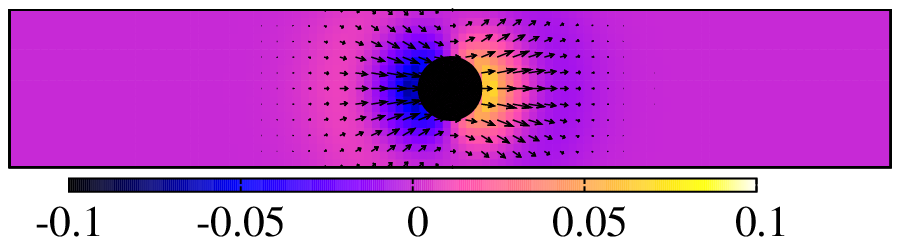}\\[0.em]
\includegraphics[width=85mm, clip]{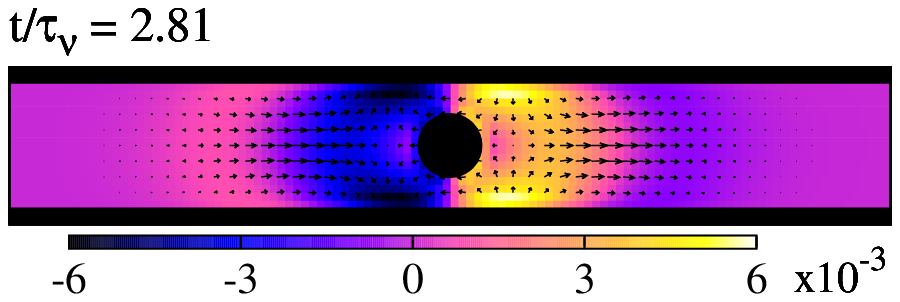}\\[-0.em]
\includegraphics[width=85mm, clip]{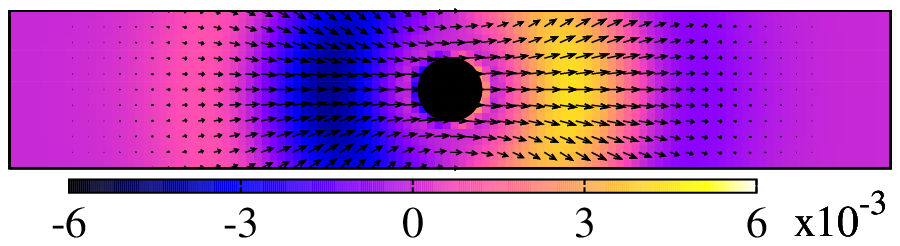}\\[0.em]
\includegraphics[width=85mm, clip]{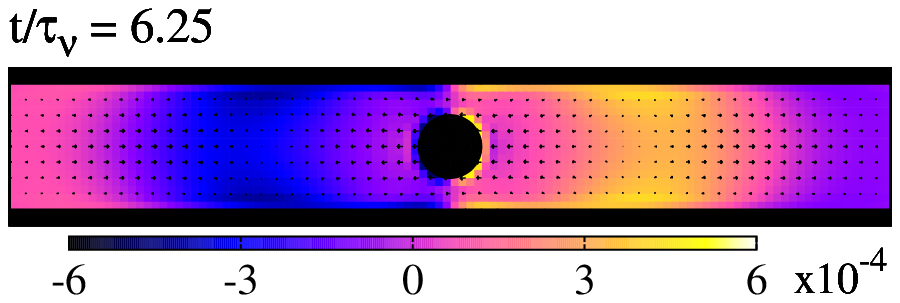}\\[-0.em]
\includegraphics[width=85mm, clip]{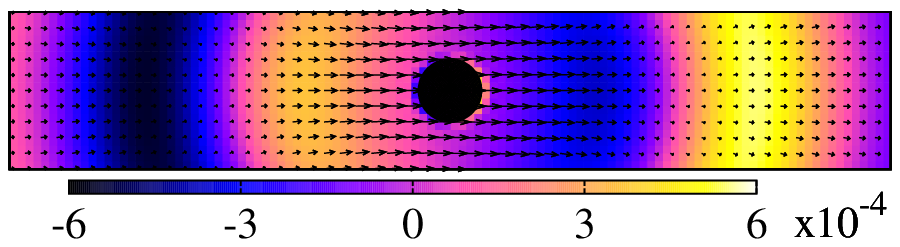}\\[-0.5em]
\caption{\label{f4-6} Temporal evolution of the velocity fields 
in a confined fluid with $H^{\ast} = 4$ (upside in each time) and in a bulk fluid (downside in each time).
The compressibility factor is $\varepsilon = 0.6$.
The walls are set on the upper and lower boundaries of the images for the confined fluids, 
and the corresponding regions are presented for the bulk fluids.
The cross-sections presented here are perpendicular to the walls and parallel to the impulsive force direction and include the particle center.
The direction of the impulsive force is to the right in the images, 
and the particle is represented by a black circle.
The divergence of the velocity field $\boldsymbol{\nabla} \cdot \boldsymbol{v}$ is described using a color scale, 
moving from negative (darker) to positive (lighter) vorticity.
The divergence is normalized by the factor of $\tau_{\nu}/{\rm Re}_p$.
}
\end{figure}

\begin{figure*}[tbp]
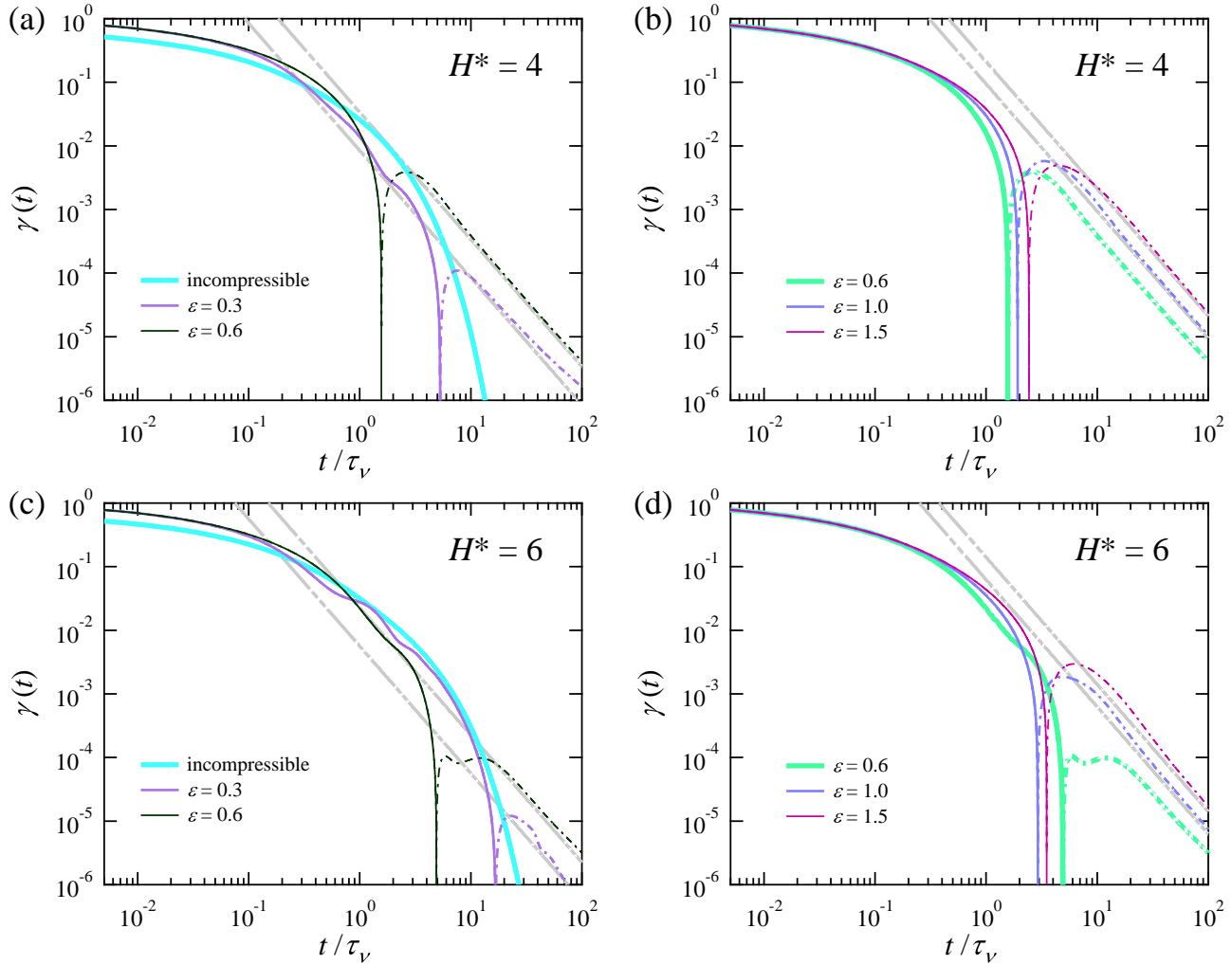

\centering
\includegraphics[width=90mm, clip]{fig4_7a.eps}\hspace{-1.0em}
\includegraphics[width=90mm, clip]{fig4_7b.eps}\\[-1.8em]
\includegraphics[width=90mm, clip]{fig4_7c.eps}\hspace{-1.0em}
\includegraphics[width=90mm, clip]{fig4_7d.eps}\\[-1.5em]
\caption{\label{f4-7} Velocity relaxation function of a particle in a compressible fluid confined between two flat walls.
The wall spacings are (a, b) $H^{\ast} = 4$ and (c, d) $H^{\ast} = 6$.
The compressibility factor has values of $\varepsilon =$ 0.3, 0.6, 1.0, and 1.5.
The simulation results are represented by solid lines (positive values) and dashed-dotted lines (negative values).
The bold dashed-two dotted lines represent the negative long-time tail $-Bt^{-2}$, as given by Eq.~(\ref{e4-3C-1}).
}
\end{figure*}

\begin{figure}[tbp]
\centering
\includegraphics[height=90mm]{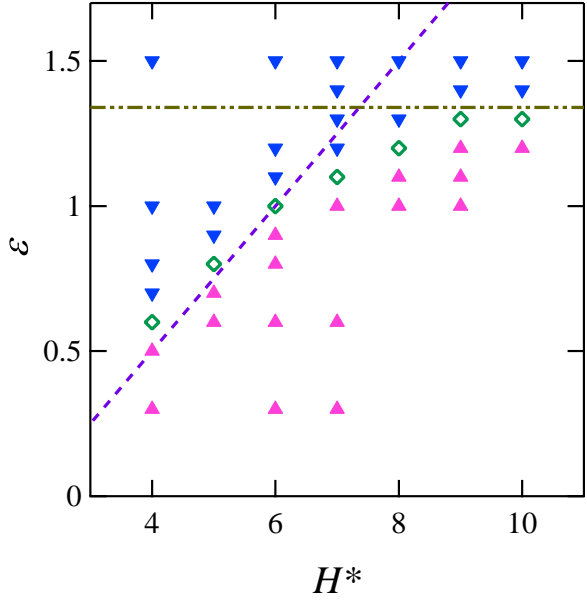}\\[-1.5em]
\caption{\label{f4-8} Classification of calculated velocity relaxation functions for various wall spacings and compressibility factors.
The time when the sign inversion of the velocity relaxation function occurs non-monotonically changes with respect to the compressibility factor.
For each wall spacing, with the increase in the compressibility factor, 
the decrease and increase of the time of the sign inversion are represented by upward and downward triangles, respectively.
The time of the sign inversion is minimized at the characteristic compressibility factor, which is represented by open diamond.
The broken line represents Eq.~(\ref{e4-3C-3}) with $\varepsilon^{\dagger} = 1$.
The dashed double-dotted line represents $\varepsilon = 1.34$.
}
\end{figure}

\begin{figure}[tbp]
\centering
\includegraphics[width=85mm, clip]{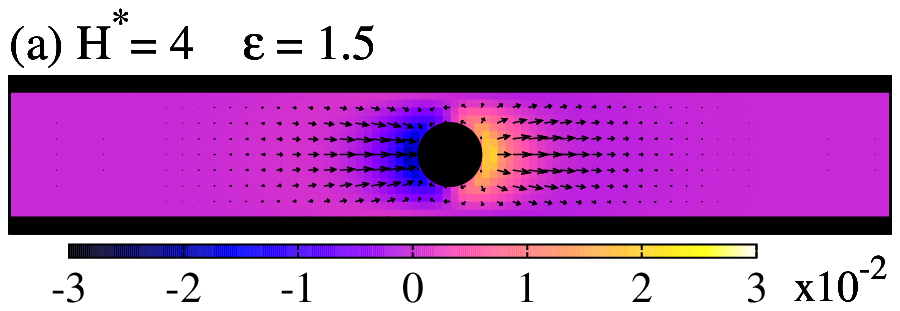}\\[0.em]
\includegraphics[width=85mm, clip]{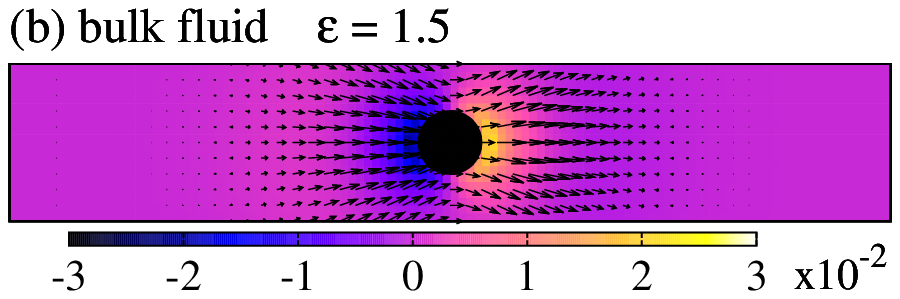}\\[-0.5em]
\caption{\label{f4-9} Velocity fields around the particle at $t/\tau_{\nu} = 2.81$ 
(a) in a confined fluid with wall spacing $H^{\ast} = 4$ and (b) in a bulk fluid.
The compressibility factor is $\varepsilon = 1.5$.
The walls are set on the upper and lower boundaries of picture for the confined fluid, 
and the corresponding region is exhibited for the bulk fluid.
The cross-sections shown here are perpendicular to the walls and parallel to the impulsive force direction with including the particle center.
The direction of the impulsive force is to the right in the pictures 
and the particle is represented by a black circle.
The divergence of the velocity field $\boldsymbol{\nabla} \cdot \boldsymbol{v}$ is described by a color scale, 
going from negative (darker) to positive (lighter) vorticity.
The divergence is normalized by the factor of $\tau_{\nu}/{\rm Re}_p$.
}
\end{figure}

In a compressible fluid, the velocity relaxation function exhibits essentially different behaviors from those in an incompressible fluid.
The simulation results of the relaxation functions are presented in Fig.~\ref{f4-5}.
Remarkable differences are the sign inversion and the subsequent negative long-time algebraic decay.
The power of the long-time decay is $t^{-2}$, and the coefficient is analytically derived as~\cite{B4-3,B4-4}
\begin{eqnarray}
\gamma(t) = -\frac{3}{8} \frac{\rho_p}{\rho_0} \frac{\varepsilon^2}{H^{\ast}} \left( \frac{\tau_{\nu}}{t} \right)^{2} \ \ \ {\rm as} \ \ t \rightarrow \infty
\label{e4-3C-1}.
\end{eqnarray}
In the present simulation results, the asymptotic approaches to Eq.~(\ref{e4-3C-1}) are accurately reproduced.
In a bulk fluid, the long-time decay is positive and proportional to $t^{-3/2}$, given by Eq.~(\ref{e4-3B-2}) as an incompressible fluid.
Although the sign inversion can occur for a large compressibility factor as $\varepsilon = 1.5$,
the relaxation function eventually becomes positive.

For the compressibility factor $\varepsilon = 0.6$, 
the temporal evolution of the velocity field in a confined fluid is compared with that in a bulk fluid in Fig.~\ref{f4-6}.
The flow accompanied by a sound wave is source-sink flow: the flow moves from the source (positive divergence region) to the sink (negative divergence region). 
The divergence of the velocity field corresponds to the fluid density deviation as
\begin{eqnarray}
\boldsymbol{\nabla} \cdot \boldsymbol{v} = -\frac{D}{Dt} \ln \frac{\rho}{\rho_0}
\label{e4-3C-2}.
\end{eqnarray}
In a bulk fluid, a sound wave propagates away from the particle; 
therefore, eventually, the effect of the sound wave on the particle motion disappears, and the particle motion is governed by shear flow accompanied by viscous diffusion.
However, in a confined fluid, a sound wave is repeatedly reflected at the walls and continuously affects the particle motion.
Multiple reflections at the walls results in the spreading of sound wave, which is known as the overdamped diffusive mode~\cite{B4-3,B4-4}.
Because viscous diffusion is prevented by the walls and the corresponding shear flow gradually disappears, 
the particle motion is eventually governed by the sound wave.
The spreading sound wave is associated with backward fluid flow and causes backtracking of the particle, 
which corresponds to the negative velocity relaxation function.

The long-time decay of the particle velocity reflects the eventual mechanism transmitting the fluid flow: viscous diffusion or sound propagation.
In a bulk fluid, the particle velocity finally decays, obeying Eq.~(\ref{e4-3B-2}), 
which corresponds to viscous diffusion of the shear flow. 
The volume of the viscous diffusion region at time $t$ is proportional to $(\nu t)^{3/2}$.
In a confined fluid, the different long-time decay Eq.~(\ref{e4-3C-1}) appears. 
This negative decay originates from the spreading sound wave, whose spreading volume at time $t$ is proportional to $H(ct)^2$.

In Fig.~\ref{f4-7}, the relaxation functions with wall spacings $H^{\ast} = 4$ and 6 for various compressibility factors are displayed.
Discrepancies of long-time decay with Eq.~(\ref{e4-3C-1}) for small compressibility factors 
are presumed to result from system size effects due to the periodic boundary conditions, 
which will appear after the time when the sound wave generated by the particle reaches the end of the system: 
$t = (L_x/2-a)/c = (L_x/2a - 1) \varepsilon \tau_{\nu}$.
The time when the sign inversion occurs changes non-monotonically with respect to the fluid compressibility.
With an increase in the compressibility factor, 
the time of the sign inversion occurs earlier when the compressibility factor is small;
however, the sign inversion occurs later when the compressibility factor is large.
In other words, there is a compressibility factor at which the time of the sign inversion is minimized, 
which we call the characteristic compressibility factor.
The characteristic compressibility factor depends on the wall spacing $H^{\ast}$;
for example, from Fig.~\ref{f4-7}, the characteristic compressibility factors are presumed to be $\varepsilon \approx 0.6$ for $H^{\ast} = 4$
and $\varepsilon \approx 1.0$ for $H^{\ast} =6$.
More detailed evaluations of the characteristic compressibility factors for various wall spacings is provided in Fig.~\ref{f4-8}.

The sign inversion of the velocity relaxation function, 
namely, the backtracking of the particle, is expected to occur after the sound wave reflected at the wall reaches the particle.
The time scale of this event is estimated by $\tau_c^{\dagger} = 2(H/2-a)/c = (H^{\ast}-2)\varepsilon \tau_{\nu}$.
Moreover, sufficient decay of the shear flow around the particle is required, 
and the time scale of the shear flow decay is $\tau_{\nu}^{\dagger}$, as given in the previous section.
Here, we define the confined compressibility factor as the ratio of the two time scales:
\begin{eqnarray}
\varepsilon^{\dagger} = \frac{\tau_c^{\dagger}}{\tau_\nu^{\dagger}} = \frac{4\varepsilon}{H^{\ast} - 2}
\label{e4-3C-3}.
\end{eqnarray}
When the confined compressibility factor is small, $\varepsilon^{\dagger} < 1$,
the spreading of the sound wave caused by reflection at the walls progresses faster than the decay of the shear flow.
For smaller confined compressibility factors,
the sound wave spreads and is weakened more rapidly; 
 therefore, further reduction of the shear flow is required to cause backtracking of the particle.
In short, when $\varepsilon^{\dagger} < 1$ is satisfied, 
backtracking of the particle occurs at a later time for a smaller confined compressibility factor.
However, when the confined compressibility factor is large, $\varepsilon^{\dagger} > 1$,
the reflected sound wave reaches the particle after sufficient decay of the shear flow; 
therefore, particle backtracking occurs at a later time for a larger confined compressibility factor.
Consequently, the characteristic compressibility factor is expected to satisfy $\varepsilon^{\dagger} \approx 1$.
In Fig.~\ref{f4-8}, the line of Eq.~(\ref{e4-3C-3}) with $\varepsilon^{\dagger} = 1$ 
almost coincides with the characteristic compressibility factors at $H^{\ast} \leq 6$; 
however, this relation fails at $H^{\ast} > 6$, 
where another mechanism of backtracking must be considered.

For the compressibility factor $\varepsilon = 1.5$, 
a comparison of the velocity fields in confined and bulk fluids at time $t/\tau_{\nu} = 2.81$ is presented in Fig.~\ref{f4-9}.
Although this time is earlier than the time $\tau_c^{\dagger}$, the relaxation function in the confined fluid is negative, as illustrated in Fig.~\ref{f4-5}(b);
therefore, the backtracking is caused by the pressure from the sound wave remaining in the vicinity of the particle. 
Backtracking by this mechanism is also observed in a bulk fluid when the compressibility factor is sufficiently large~\cite{B4-19,B4-20}, 
and the condition for which the backtracking occurs is estimated as $\varepsilon \geq 1.34$ from the analytical solution~\cite{B4-18,B4-19}.
There is only a slight difference in the divergence of the velocity field between the confined and bulk fluids;
however, in a confined fluid, because shear flow is diminished by the walls, 
backtracking by a sound wave can occur at an earlier time than in a bulk fluid, as demonstrated in Fig.~\ref{f4-5}(b).
Considering this mechanism of particle backtracking, 
the characteristic compressibility factor will eventually converges to $\varepsilon = 1.34$ with an increase in the wall spacing $H^{\ast}$.
The successive change in the dependence of the characteristic compressibility factor on the wall spacing, 
namely, from $\varepsilon^{\dagger} = 1$ to $\varepsilon = 1.34$, is observed in Fig.~\ref{f4-8}.
The wall spacing at which this crossover occurs is estimated as $H^{\ast} \approx 7.4$.

\section{Conclusion}

We investigated the dynamics of a single particle in a fluid confined by two parallel plane walls 
using SPM to perform direct numerical simulations.
In particular, we calculated the velocity relaxation of the particle after an impulsive force was added in the direction parallel to the walls. 
The velocity relaxation function corresponds to the velocity autocorrelation function in a fluctuating system.

In an incompressible fluid, the relaxation function decayed more rapidly in a confined fluid than in a bulk fluid; 
the long-time decay was exponential differently from the power law $t^{-3/2}$ observed in a bulk fluid.
The rapid decay reflects the loss of shear flow due to the hindrance of viscous diffusion by the walls.
Therefore, the time when the relaxation function in a confined fluid falls below that in a bulk fluid corresponds to 
the time scale of viscous diffusion over the distance between the particle surface and the walls, $\tau_{\nu}^{\dagger}$.

In a compressible fluid, sign inversion and subsequent negative long-time decay obeying $t^{-2}$ 
were observed in the velocity relaxation functions.
A corresponding spreading sound wave arising from the multiple reflections at the walls was also observed.
The time of the sign inversion changed non-monotonically with respect to the compressibility factor; 
that is, the time of the sign inversion was minimized at the characteristic compressibility factor.
The characteristic compressibility factor satisfies $\varepsilon^{\dagger} = 1$ when the wall spacing is small
and is given by $\varepsilon = 1.34$ when the wall spacing is large, 
where the confined compressibility factor $\varepsilon^{\dagger}$ is defined by Eq.~(\ref{e4-3C-3}).
The crossover of the dependence of the characteristic compressibility factor on the wall spacing is estimated to be $H^{\ast} \approx 7.4$.
The backtracking of the particle is caused by the spreading sound wave arising from reflections at the walls 
when the characteristic compressibility factor satisfies $\varepsilon^{\dagger} = 1$ 
and is caused by the remaining sound wave in the vicinity of the particle 
when the characteristic compressibility factor is given by $\varepsilon = 1.34$.
Backtracking via the latter mechanism can also occur in a high compressible bulk fluid.

\begin{acknowledgments}
This work was supported by KAKENHI 23244087 and 
the JSPS Core-to-Core Program ``International research network for non-equilibrium dynamics of soft matter.''
\end{acknowledgments}

\nocite{*}

\bibliography{apssamp}

\end{document}